\documentstyle[preprint,aps]{revtex}
\begin{document}
\draft
\preprint{RU9466}
\title{Spin-rotor Interpretation of Identical Bands and Quantized
Alignment in Superdeformed A $\approx$ 190 Nuclei}
\author{J. A. Cizewski}
\address{Department of Physics and Astronomy, Rutgers University,
New Brunswick, NJ 08903}
\author{R. Bijker}
\address{Instituto de Ciencias Nucleares, U.N.A.M.,
Apartado Postal 70-543, 04510 M\'exico, D.F., M\'exico}
\author{J. A. Becker, M. J. Brinkman\cite{byline}, E. A. Henry}
\address{Lawrence Livermore National Laboratory, Livermore, CA 94550 }
\author{F. S. Stephens, M. A. Deleplanque, R. M. Diamond}
\address{Lawrence Berkeley Laboratory, Berkeley, CA 94750 }
\date{\today}
\maketitle
\begin{abstract}
The ``identical'' bands in superdeformed mercury, thallium, and
lead nuclei are interpreted as examples of orbital angular
momentum rotors with the weak spin-orbit coupling of pseudo-$SU(3)$
symmetries and supersymmetries.
\end{abstract}
\narrowtext

\section{INTRODUCTION}

	Considerable excitement in nuclear structure science has
been generated by the observation of cascades of $\gamma$-ray
transitions in superdeformed (SD) rotational bands with
``identical'' energies.  The initial observation~\cite{1} was in
pairs of A $\approx$ 150 nuclei ($^{151}$Tb$^{\star}$-$^{152}$Dy) and
($^{150}$Gd$^{\star}$-$^{151}$Tb) where
the $\gamma$-ray energies in the pairs were identical within 1-3 keV.
This was followed shortly by the observation~\cite{2}  of excited
superdeformed bands in $^{194}$Hg with $\gamma$-ray energies related to
those of the only known SD band in $^{192}$Hg:  one of these bands
has $\gamma$-ray energies identical to those in SD $^{192}$Hg, the other
has energies which occur at the arithmetic mean or ``midpoint''
values.  In the following ``identical'' will refer to $\gamma$-ray
transition energies which are simply related to those in a
reference band, and will include ``quarter points'', as well as
midpoint and equal energies.  This paper will propose a new
rotational coupling scheme in which the observed pattern of
$\gamma$-ray transition energies is predicted.  Preliminary, but
incomplete, reports of this work have been previously
presented~\cite{3}.

	Since the initial observations and with the advent of
the largest arrays, Eurogam and Gammasphere, a large number
of SD bands have been identified and the earlier candidates
have been more firmly established and extended to both lower
and higher $\gamma$-ray energies. The data which we shall discuss
are related to $^{192}$Hg~\cite{4} and $^{193}$Tl~\cite{5} SD reference
configurations. Related to the $^{192}$Hg reference are the yrast
SD band in $^{194}$Pb~\cite{6}, $^{191}$Hg$^{\star}$ SD-2 and
3~\cite{7}, $^{193}$Hg SD-3 and 4~\cite{8},
and $^{194}$Hg$^{\star}$ SD-2 and 3~\cite{9}. Related to the
$^{193}$Tl reference are
4 SD bands in $^{194}$Tl (SD 1-4)~\cite{10} and 2 SD bands in
$^{194}$Pb$^{\star}$ (SD-2a and 2b)~\cite{11}. The available data have
recently been compiled and we have adopted the energies and nomenclature
of Ref.~\cite{12}, supplemented with the most recent available data
from Refs.~\cite{9,10,11}.  We adopted the energies and nomenclature
of the evaluated date reported in ref. ~\cite{12} because some of the
bands have been observed by several groups, all with their own system
to name the bands.  In ref. ~\cite{12} the bands are uniformly
labled by an arabic number, with the lowest being the first,
usually strongest, band observed.  Adjacent numbers could be
signature partners, but no such assumption of physical properties
are proposed in ref. ~\cite{12}.

One of the main differences between the $A \approx 150$ and
$A \approx 190$ SD excitations is that the $A \approx 190$ SD
$\gamma$-ray cascades extend to
relatively low $E_{\gamma}$, typically $<250$ keV for the data in
Figs.~1-4.  The regular behavior of the SD cascades and the low
transition energies imply that the cascades extend down to
low angular momentum and suggest that reliable spin
assignments~\cite{13}  can be made through comparison with the
predictions of the quantum rotor.  For the bands displayed in
Figs.~1-4, the extracted $J_f$ values are within $0.1 \hbar$ of integer
(for even-even) or half-integer (for odd-$A$) values.
Uncertainties in $J_f$ values are larger for the odd-odd cases
and the weak $^{194}$Pb$^{\star}$ SD bands.

	There are two new aspects of these SD excitations which
are not expected.  First, the $\gamma$-ray energies in neighboring
nuclei are directly related to those of $^{192}$Hg  or $^{193}$Tl.  This
means that these rotational structures have identical moments
of inertia, in itself an unexpected result.  The second new
and unexpected result is the value of the difference in
angular momenta between one SD band and the reference, which
was summarized in Fig.~3 of Ref.~2b.  This additional angular
momentum, or alignment, sets in at moderate $\gamma$-ray energies and
saturates at $1.00(4) \hbar$ for the bands displayed in Figs.~1 and~2,
and is very close to $1 \hbar$ or $0 \hbar$ for four $^{194}$Tl
SD bands with respect to $^{193}$Tl.
This observation of $1 \hbar$ of alignment was not
expected, and has led to intense scrutiny of the methods used
to extract the spins. We stand by our spin assignments; a
full discussion is given in Ref.~\cite{13}.

\section{COUPLING SCHEMES WITH PSEUDO ORBITAL ANGULAR MOMENTA}

	The observation of identical bands and integer alignment
at moderate $\gamma$-ray energies suggests these rotational
excitations are dominated by an integer angular momentum, L,
with a contribution from the (pseudo-) orbital angular momenta of the
valence fermions, and only a small (pseudo-) spin-orbit interaction.
While not a new coupling scheme, it is not the usual one for
heavy nuclei, in which the single-particle structure of
nuclei is assumed to be governed by the \underline{total} angular
momentum, $\vec{j}=\vec{l}+\vec{s}$, rather than $l$ and $s$, separately.
For example, in the 50-82 nuclear shell, the orbitals are $g_{7/2}$,
$d_{5/2}$, $h_{11/2}$, $d_{3/2}$, and $s_{1/2}$; the $g_{9/2}$ orbital
is below the 50
shell gap, and the $h_{11/2}$ negative-parity orbital has come
down from the $N=5$ shell.  The normal, in this case positive
parity, orbitals in the 50-82 shell have $j=7/2$, 5/2, 3/2, 1/2;
these are exactly the $j$-values for the orbitals in a (pseudo)
$N=3$ shell.  In many cases a pseudo-harmonic oscillator~\cite{14},
or pseudo-$SU(3)$ scheme for deformed nuclei, is an appropriate
framework in which to discuss nuclear excitations, and it can
provide a good description of the observables.  We are
suggesting that a coupling dominated by the total \underline{orbital}
angular momentum, with relatively little spin-orbit
splitting, as is the case for the pseudo-harmonic oscillator,
can reproduce the observed pattern of identical SD bands in
A $\approx$ 190 nuclei.  In the following we shall discuss two
different coupling schemes in which the integer alignments can be
accommodated.

\subsection{Strong Coupling between the Spins}

	We consider a core-particle model in which
the core angular momentum $R$ is coupled with the
pseudo-orbital part $\tilde{L}$ of the nucleons outside the core
to $\vec{L}=\vec{R}+\vec{\tilde{L}}$,
which is subsequently coupled with the spin part $S$ to total
angular momentum $\vec{J}=\vec{L}+\vec{S}$.
In the first coupling scheme, in which we assume
a strong coupling between the spin parts of the angular momenta
of the nucleons outside the core, the rotational hamiltonian
for identical SD excitations is \cite{15,16}
\begin{eqnarray}
H &=& a \, \vec{S} \cdot \vec{S} + b \, \vec{L} \cdot \vec{L}
+ c \, \vec{L} \cdot \vec{S} ~.
\end{eqnarray}
The eigenvalues and $\gamma$-ray transition energies of this
hamiltonian can be written as
\begin{eqnarray}
E \;=\; A_1 \, S(S+1) + B_1 \, L(L+1) + C_1 \, J(J+1) ~,
\nonumber \\
E_{\gamma} (J+1 \rightarrow J-1) \;=\; B_1 \, (4L+2) + C_1 \, (4J+2) ~,
\end{eqnarray}
with parameters $A_1=a-c/2$, $B_1=b-c/2$ and $C_1=c/2$.
The $\gamma$-ray energies depend only on
$B_1$ and $C_1$, since the $A_1$ term only contributes to the band-head
energy.  This formulation has been proposed previously to
study pseudo-spin symmetries in superdeformed nuclei~\cite{16}.

	For a one-fermion ($N_F=1$) nucleus, $S=1/2$ and $J=L \pm 1/2$.
For a two-fermion ($N_F=2$) configuration, $S=0$ or $S=1$; since
$\vec{J}=\vec{L}+\vec{S}$, a plethora of $J$ values arises from the
vector addition.
Generic spectra for $N_F=1$ and $N_F=2$ systems are illustrated in
Figs.~5 and~6, respectively. For the $N_F=1$ case three types of
spectra are expected.  The first two in Figs.~5a and~5b,
respectively, are decoupled structures, which in a more
traditional framework are bands with decoupling parameters
$+1$ and $-1$. In both of these cases only one $\gamma$-ray cascade
would be observed, with $\gamma$-ray energies either identical to or
at midpoint values to those of the even-even reference, when
$C_1=0$.  This type of spin-rotor could explain many of the
identical bands observed in the $A \approx 150$ region, and in
particular, the $^{151}$Tb$^{\star}$-$^{152}$Dy pair of SD
bands~\cite{1}.  However,
the ``identical'' bands in $^{191,193}$Hg compared to $^{192}$Hg  are not
examples of decoupled structures, but rather can be
understood as examples of the generic spectrum of Fig.~5c.

For the $N_F=1$, $J=L+1/2$ (Fig.~5c) and $N_F=2$, $S=1$, $J=L+1$
(Fig.~6c) cases, the transition energies are given by:
\begin{eqnarray}
\begin{array}{lll}
N_F=0 & \hspace{1cm}  E_{\gamma} = (B_1+C_1)(4J+2) & \hspace{1cm}   J=L ~, \\
N_F=1 & \hspace{1cm}  E_{\gamma} = B_1(4J) + C_1(4J+2) & \hspace{1cm}  J=L+1/2
{}~, \\
N_F=2 & \hspace{1cm}  E_{\gamma} = B_1(4J-2) + C_1(4J+2) & \hspace{1cm}  J=L+1
{}~.
\end{array}
\end{eqnarray}
Although the spectra in Fig.~5 were generated with $C_1=0$, the
$\gamma$-ray energies can depend on $C_1$ without breaking the symmetry;
it is the $C_1$ term which is the additional ingredient required
to reproduce the observed alignments.  When $B_1=-2C_1$ the
transition energies in Eq.~3 become
\begin{eqnarray}
\begin{array}{lll}
N_F=0 \hspace{1cm} & E_{\gamma} = B_1(2J+1) & \hspace{1cm} J=L ~,\\
N_F=1 \hspace{1cm} & E_{\gamma} = B_1(2J-1) & \hspace{1cm} J=L+1/2 ~,\\
N_F=2 \hspace{1cm} & E_{\gamma} = B_1(2J-3) & \hspace{1cm} J=L+1 ~.
\end{array}
\end{eqnarray}

In Fig.~1b we compare the $\gamma$-ray transitions in $^{193}$Hg SD-3 and
4 bands to the expectations of a spin-rotor, Eq. 4.  With
this choice of parameters the observed alignment $i=1 \hbar$ can be
reproduced, and the same value of $B_1$ is used for the reference
and one-fermion bands.  The same quality of agreement would
have been observed if we had chosen to compare the data for
$^{191}$Hg SD-2,3 with the predictions of Eq.~4.

	For the $N_F=2$ case there are a large number of generic
spectra.  The simplest case occurs for $S=0$ and the same
orbital angular momenta, $L$, for the reference and $N_F=2$
configuration.  This gives identical $E_{\gamma}$ values, exactly what
is observed for $^{192}$Hg  and $^{194}$Pb, where the $\gamma$-ray
energies~\cite{6} are on average within 1 keV for 10 transitions.
In addition to $S=0$, $S=1$ is allowed for the two-fermion system.
That two-particle excitations in which the spins are aligned
(with $S=1$) could be important is in accord with the earlier
suggestion~\cite{10}  that triplet pairing could be used to
understand the observed $i=1 \hbar$ alignment.  Since two excited
bands are observed in $^{194}$Hg, a strongly coupled spectrum is
suggested with $S=1$ and $J=L+1$, as displayed in Fig.~6c.  With
Eq.~4 and the same parameters used to fit the odd-$A$ spectrum,
this coupling scheme gives $i=2\hbar$, as displayed in Fig.~2b,
but in contrast to the data.  The hamiltonian of Eq.~1 cannot
then be used to fit simultaneously the data for the single
neutron and two neutron excitations, if we require the same
parameters for all nuclei in a multiplet.  Rather, when $B_1=-2C_1$
one gets $i=1 \hbar$ for $N_F=1$ systems and $i=2\hbar$ for $N_F=2$;
or when $C_1=0$ one gets $i=\hbar/2$ for $N_F=1$ and $i=1\hbar$ for
$N_F=2$ systems.

In $^{194}$Tl six SD bands have been identified~\cite{10}; two of these
have zero alignment with respect to the $^{193}$Tl reference, and two
have $i \approx 1\hbar$, as shown in Fig.~3a.  $^{194}$Tl is again a
two-fermion system, for which we propose $S=1$.
In Fig.~3b we present the expectations for the reference
($N_F=0$), $N_F=1$ ($^{193}$Tl) and $N_F=2$, $S=1$ bands.
The $\gamma$-ray transitions in Eq.~4 give $N_F=2$ bands with
$1\hbar$ of alignment with respect to the one-fermion core, which
reproduces the pattern for SD-1 and 2 bands in $^{194}$Tl.  In
contrast, the SD-3 and 4 bands in $^{194}$Tl have zero alignment
with respect to $^{193}$Tl, or $1\hbar$ of alignment with respect to an
even-even core.  As was the case for the two-neutron bands in
$^{194}$Hg, the coupling scheme of Eq.~2, with the same parameters
for the $N_F=1$ and $N_F=2$ nuclei, cannot reproduce the data for
SD-3 and 4 in $^{194}$Tl.

The coupling scheme of Eq.~2 assumes that the spins of
the two fermions couple strongly to $S=1$.  For identical fermions,
the Pauli principle requires that the total wave function is
antisymmetric.  Therefore, the relative orbital angular
momentum of the two identical fermions must be odd.  There is
also no guarantee that both excited fermions in $^{194}$Hg will
come from the same oscillator shell as the single fermion in
$^{193}$Hg, for example, and clearly the odd-neutron and odd-proton
in $^{194}$Tl can be expected to have very different
configurations.  Therefore, it may not always be correct to
assume the strong coupling of the spins of the two fermions.

The level diagram for single-particle configurations at
large deformations is a complicated mixture of orbitals from
many shells.  For example, the neutron orbitals for Hg nuclei
in their ground states are mostly from the $N=5$ oscillator
shell; at large deformations one also finds many orbitals
from the $N=6$ shell, as well as ``intruder'' $j_{15/2}$ configurations
from the $N=7$ shell.  Only the isolated high-$j$ $N=7$ orbitals
need be considered as outside of the framework of a symmetry.
Therefore, the two-fermion system can either have two
particles in the same shell, or each fermion can come from an
orbital from a different shell.

\subsection{Weak Coupling between the Spins}

Again we consider a core-particle model in which the core angular
momentum $R$ is coupled with the pseudo-orbital part $\tilde{L}$ of
the nucleons outside the core to $\vec{L}=\vec{R}+\vec{\tilde{L}}$.
In this case we anticipate
that the two fermions are from different oscillator (or
pseudo-oscillator) shells.  Therefore, the spins are not coupled
together, but rather $\vec{J}_1=\vec{L}+\vec{S}_1$ is the
angular momentum involving the spin of one of the fermions,
and $\vec{J}=\vec{J}_1+\vec{S}_2$ is the total angular momentum.
In this model the excitation spectrum and $\gamma$-ray transition
energies are given by
\begin{eqnarray}
E \;=\; B_2 \, L(L+1) + C_2 \, J_1(J_1+1) + D_2 \, J(J+1) ~,
\hspace{1cm}
\nonumber\\
E_{\gamma} (J+1 \rightarrow J-1) \;=\; B_2 \, (4L+2)
+ C_2 \, (4J_1+2) + D_2 \, (4J+2) ~.
\end{eqnarray}

Again, there will be a large variety of bands arising from the
different ways the angular momenta can be coupled to total $J$.
For the generic spectra illustrated in Fig.~7, with $J=J_1+1/2$
for the $N_F=2$ nucleus, the transition energies are
\begin{eqnarray}
\begin{array}{lll}
N_F=0 \hspace{1cm} & E_{\gamma} = (B_2+C_2+D_2) \, (4J+2)
& \hspace{1cm} J=J_1=L ~,\\
N_F=1 \hspace{1cm} & E_{\gamma} = B_2 \, (4J) + (C_2+D_2) \, (4J+2)
& \hspace{1cm} J=J_1=L+1/2 ~,\\
N_F=2 \hspace{1cm} & E_{\gamma} = B_2 \, (4J-2) + C_2 \, (4J)
+ D_2 \, (4J+2) & \hspace{1cm} J=J_1+1/2=L+1 ~.
\end{array}
\end{eqnarray}
Alignment $i=1\hbar$ can be obtained in the $N_F=1$ \underline{and}
$N_F=2$ systems when $C_2=-B_2$ and $2D_2=B_2$. For this case Eq.~6
becomes
\begin{eqnarray}
\begin{array}{lll}
N_F=0 \hspace{1cm} & E_{\gamma}  = B_2 \, (2J+1)
& \hspace{1cm} J=J_1=L ~,\\
N_F=1 \hspace{1cm} & E_{\gamma}  = B_2 \, (2J-1)
& \hspace{1cm} J=J_1=L+1/2 ~,\\
N_F=2 \hspace{1cm} & E_{\gamma}  = B_2 \, (2J-1)
& \hspace{1cm} J=J_1+1/2=L+1 ~.
\end{array}
\end{eqnarray}

A comparison between experiment and the predictions from
Eq.~7 for $^{192}$Hg - $^{194}$Hg(SD-2,3) and $^{193}$Tl -
$^{194}$Tl(SD-3,4) are
shown in Figs.~2 and 3, respectively.  The data for $^{194}$Tl
actually require two different coupling schemes for this odd-odd
nucleus:  (i) the orbital angular momenta of both
fermions are strongly coupled, and their spins couple to $S=1$
(Eq.~2); (ii) the spin of the second fermion is weakly coupled
to the total angular momentum of the first fermion (Eq.~5).
This should not be unexpected since the odd proton is most
likely in an $i_{13/2}$ orbital~\cite{5}, while the odd neutron
could be in either an $N=6$ or $N=5$ orbital, which have very different
radial overlaps with respect to the proton orbit.  Different
predictions come from these two coupling schemes. Equation~2
arises when both fermions are in the same shell, and a
positive-parity band will result, while that of Eq.~5 arises
when the fermions are in orbitals from different shells, so
that it is quite likely that the parity of the SD band will
be negative.  A measure of the parities of these SD
excitations could further test these predictions.

The cases on which we have focused were identified with
the previous generation of large arrays of $\gamma$-ray detectors.
In the past two years there has been an explosion of new data with
the first results from the larger arrays, Eurogam and
Gammasphere.  One of these results was the identification~\cite{11}
of SD excited bands in $^{194}$Pb.  As displayed in Fig.~4 the
$\gamma$-ray energies and spins of these bands indicate zero alignment
with respect to $^{193}$Tl.  This is another example of the
coupling scheme of Eq.~5, which indicates that the two
excited protons are probably in orbitals from different major
shells.

In the present analysis we have not attempted to
superimpose the  predictions on the data.  The main reason:
while the moments of inertia are identical for these nuclei,
they are not constant as a function of spin.  Rather, the
dynamical moments of inertia increase by $\approx$50\% over the
measured range of $\gamma$-ray energies.  This could be reproduced by
allowing $B_1$ or $B_2$, the only free parameters, to have a
dependence on spin.

\section{DISCUSSION}

The spin-rotor interpretation of the identical bands and
quantized alignment is included in a number of nuclear
structure models which involve good rotors and the pseudo-harmonic
oscillator.  For example, the identical bands in the
$A \approx 150$ and 190 regions have been proposed as examples of a
dynamical supersymmetry~\cite{15,17}.   For a boson-fermion deformed
or $SU(3)$ symmetry, the eigenvalue equations in both Eqs.~2 and~5
can be appropriate.  A supersymmetry is a valid description
when the same parameters are used for the even core and the
one fermion, or two-fermion, systems. The observation of identical
behavior in $^{194}$Hg$^{\star}$ and $^{192}$Hg is then the first
candidate for a \underline{two}-fermion dynamical supersymmetry.  In
addition, the spin-rotor is also part of the fermion pseudo-$SU(3)$
framework~\cite{14}, where again Eq.~2 is valid~\cite{16} for the
one-fermion system and can be extended to two-fermion
excitations.  However, the coupling scheme of Eq.~5 does not
naturally occur in this latter framework~\cite{18}.

The spin-rotor interpretation of the identical SD bands
is based on the assumption that the additional particle(s)
are in orbitals that can be assigned either harmonic
oscillator or pseudo-harmonic oscillator quantum numbers,
although it has been recognized that the asymptotic pseudo-harmonic
oscillator behavior often better explains the spectroscopic
properties at finite deformation.  It is accepted~\cite{5} that
the odd-proton in $^{193}$Tl SD bands is in an $i_{13/2}$ orbital,
which is separated from other $N=6$ orbitals, and therefore these
SD bands are not expected to be simply related to the $N_F=0$
reference, $^{192}$Hg.  The SD-3 and 4 bands in $^{193}$Hg have
been suggested~\cite{8} to come
from the $i_{13/2}$ extruder orbital.  While such a configuration
is outside of the pseudo-harmonic oscillator framework, these
SD bands in $^{193}$Hg are observed to be simply related to the
$N_F=0$ reference, $^{192}$Hg SD. The exact ordering of orbitals at
these large deformations is sensitive to the parameters used
for the $\ell^2$ and spin-orbit terms in the single-particle
potential. While most studies have assumed the ordering of
single-particle orbitals given in Ref.~9b, no model
independent measure of the configurations involved in the SD
bands exists.  In contrast to the calculations in Ref.~9b,
Nilsson calculations using parameters by $\AA$berg~\cite{19}  predict
the 7/2$^-$[514] and 5/2$^-$[512] orbitals to be close in energy to
the 9/2$^+$[624] orbital assigned to SD-3,4 in $^{193}$Hg in Ref.~8.
These $N=5$ orbitals are $N=4$ pseudo-spin partners and within
the present framework.  Given that such critical properties
as spin, parity, and excitation energy have not been
determined, we shall have to wait for more definitive
measures of the spectroscopic properties of these SD bands to
test the microscopic basis of the spin-rotor predictions.

\section{CONCLUSIONS}

In summary, we are able to understand both the $\gamma$-ray
energies and extracted alignments of a large number of the
superdeformed rotational bands in mercury, thallium, and lead
nuclei as examples of quantum rotors in which an \underline{orbital}
angular momentum plays the dominant role, with only a weak
dependence on the total angular momentum, which arises from a
relatively small spin-orbit interaction.  This is a new
coupling scheme for heavy nuclei.  Traditionally, the \underline{total}
angular momenta carried by the particles dominates the
coupling, because of the strong spin-orbit interaction.  The
spin-rotor scheme arises naturally in models which involve
pseudo orbital angular momenta, for example, pseudo-$SU(3)$
models of fermions, or bosons and fermions.  We have shown
that this coupling scheme is not only valid for one-fermion
systems, but also for two-fermion excitations, and the
$^{192}$Hg--$^{193}$Hg--$^{194}$Hg$^{\star}$ multiplet,
and possibly $^{192}$Hg--$^{193}$Tl--$^{194}$Pb$^{\star}$, could
be the first examples of a multi-fermion supersymmetry.

With the advent of the new, large arrays of high-resolution Ge
detectors, such as Eurogam and Gammasphere,
there has been an explosion in the number of superdeformed
rotational bands which have been identified~\cite{20},  and in a
large number of these new cases identical bands have been
observed.  We look forward to these new results and, in
particular, the confirmation of spin and parity assignments
which will be possible when definitive links between
superdeformed and normal excitations have been identified.

\acknowledgments

We would like to thank F. Azaiez and J. R. Hughes for
providing experimental results prior to publication.  One of
us (JAC) would like to thank Professors F. Iachello, A.
Arima, and J. P. Draayer for invaluable discussions.  This
work was supported in part by the National Science Foundation
(Rutgers), by CONACyT, M\'exico under project
400340-5-3401E, DGAPA-UNAM under project IN105194, and the U.S.
Department of Energy under contract numbers W-7405-ENG-48
(LLNL) and DE-AC03-76SF-00098 (LBL).

\begin{figure}
\caption{Average angular momentum in units of $\hbar$ as a
function of $E_{\gamma}$ for
(a) SD-3 and 4 in $^{193}$Hg (circles) compared to the $^{192}$Hg
reference(closed triangles);  (b) the spin-rotor (Eq.~4) with
$N_F=1$, (circles) compared to the $N_F=0$, $S=0$ reference (closed
triangles).  The parameter $B_1=10$ keV.  Data are taken from
Refs.~4 and~8.  }  The average angular momentum is the value of
J for the $\gamma$-ray transition between levels with J+1 and J-1.
\end{figure}

\begin{figure}
\caption{Average angular momentum in units of $\hbar$ as a
function of $E_{\gamma}$ for
(a) SD-2 and 3 in $^{194}$Hg (circles) compared to the $^{192}$Hg
reference (closed triangles);  (b) the spin-rotor (Eq.~4) with
$N_F=2$, $S=1$ (circles) compared to the $N_F=0$, $S=0$ reference
(closed triangles), and the $N_F=2$, $J=J_1+1/2$ spectrum (Eq.~7)
(squares).  The parameters $B_1=B_2=10$ keV.  Data are taken
from Refs.~4 and~9.  }
\end{figure}

\begin{figure}
\caption{Average angular momentum in units of $\hbar$ as a
function of $E_{\gamma}$  for
(a) SD bands 1-4 in $^{194}$Tl (circles and squares) compared to
the $^{193}$Tl reference (triangles); (b) the spin-rotor (Eq.~4)
with $N_F=2$, $S=1$ (circles) compared to the $N_F=0$, $S=0$ (closed
triangles) and $N_F=1$ (small diamonds) references, and the $N_F=2$,
$J=J_1+1/2$ spectrum (Eq.~7) (squares). The parameters $B_1=B_2=10$
keV.  Data are taken from Refs.~5 and~10.}
\end{figure}

\begin{figure}
\caption{Average angular momentum in units of $\hbar$ as a
function of $E_{\gamma}$ for
(a) SD-2a and 2b in $^{194}$Pb (circles) compared to the $^{193}$Tl
reference (triangles); (b) the $N_F=2$, $J=J_1+1/2$ spectrum (Eq.~7)
(squares) compared to the $N_F=0$, $S=0$ (closed triangles) and
$N_F=1$ (small diamonds) references.  The parameter $B_2=10$ keV.
Data are taken from Refs.~5 and~11.}
\end{figure}

\begin{figure}
\caption{Generic spectra for $N_F=1$, $S=1/2$ spin-rotors
(Eq.~2) with $C_1=0$.  Type c corresponds to signature partner
bands and are compared to data in odd-$A$ candidates.  The
left-hand spectra are the references.}
\end{figure}

\begin{figure}
\caption{Generic spectra for $N_F=2$, $S=1$, spin-rotors (Eq.~2)
with $C_1=0$.  Type c corresponds to signature partner bands and
are compared to data in two-fermion candidates.  The left-hand spectra
are the references.}
\end{figure}

\begin{figure}
\caption{Generic spectra for $N_F=1$ and $N_F=2$, $J=J_1+1/2$
spin-rotors (Eq.~5) with $C_2=D_2=0$. The left-hand spectrum is the
reference.}
\end{figure}

\end{document}